\documentclass[prd,aps,floats,twocolumn]{revtex4}
\usepackage{commath}
\pdfoutput=1
\usepackage[usenames, dvipsnames]{color}
 \usepackage{graphicx, array}
\usepackage{graphics,epsfig}
 \usepackage{amssymb}
 \usepackage{amsmath}
 \usepackage{ulem}
 \usepackage{multirow}
  \usepackage{subfigure}
\usepackage{bm}
\usepackage{txfonts}
\usepackage{cancel}
\newcolumntype{C}[1]{>{\centering\arraybackslash}m{#1}}

 \def\be{\begin{equation}}
\def\ee{\end{equation}}
 \def\bi{\begin{itemize}}
 \def\ei{\end{itemize}}
  \def\ben{\begin{enumerate}}
\def\een{\end{enumerate}}
  \def\bt{\begin{tabular}}
\def\et{\end{tabular}}
\def\bc{\begin{center}}
\def\ec{\end{center}}

\def\deltam{{\delta}}

\def\bea{\begin{eqnarray}}
\def\eea{\end{eqnarray}}

\def\ba{\begin{eqnarray}}
\def\ea{\end{eqnarray}}

\begin{document}

\input{epsf}

\title{Beyond $\delta$: Tailoring marked statistics to reveal modified gravity}
\author {Georgios Valogiannis and Rachel Bean.}
\affiliation{Department of Astronomy, Cornell University, Ithaca, New York 14853, USA.}
\label{firstpage}

\begin{abstract}

Models that seek to explain cosmic acceleration through modifications to General Relativity (GR) evade stringent Solar System constraints through a restoring, screening mechanism. Down-weighting the high density, screened regions in favor of the low density, unscreened ones offers the potential to enhance the amount of information carried in such modified gravity models.  

In this work, we assess the performance of a new ``marked" transformation and perform a systematic comparison with the clipping and logarithmic transformations, in the context of $\Lambda$CDM and the symmetron and $f(R)$ modified gravity models. Performance is measured in terms of the fractional boost in the Fisher information and the signal-to-noise ratio (SNR) for these models relative to the statistics derived from the standard density distribution. We find that all three statistics provide improved Fisher boosts over the basic density statistics. The model parameters for the ``marked" and clipped transformation that best enhance signals and the Fisher boosts are determined. 
We also show that the mark is useful both as a Fourier and  real space transformation; a marked correlation function also enhances the SNR relative to the standard correlation function, and can on mildly non-linear scales show a significant difference between the $\Lambda$CDM and the modified gravity models.

Our results demonstrate how a series of simple analytical transformations could dramatically increase the predicted information extracted on deviations from GR, from large-scale surveys, and give the prospect for a potential detection much more feasible. 
\end{abstract}

\maketitle

\section{Introduction}
\label{sec:intro}
Theories that invoke large-scale modifications to General Relativity (GR), the so-called Modified Gravity (MG) theories \cite{CLIFTON20121}, are popular theoretical attempts to explain the recent accelerative phase of the universe, as observed by a wide range of observational probes \cite{Perlmutter:1998np,Riess:2004nr,Eisenstein:2005su,Percival:2007yw,Percival:2009xn,Kazin:2014qga,Spergel:2013tha,Ade:2013zuv,Ade:2015xua}. The simplest $\Lambda$CDM cosmological scenario, that produces acceleration within the framework of GR through a cosmological constant, $\Lambda$, is a great fit of the data and thus widely accepted, but suffers from undesirable fine-tuning problems \cite{Weinberg:1988cp}, that consequently motivated  the consideration of a range of alternatives. Any attempt to provide a robust theoretical explanation of cosmic acceleration through modifications to gravity, however, should satisfy the stringent experimental constraints of GR in the vicinity of the solar system \cite{Will:2005va}, which would in principle be violated by an additional degree of freedom \cite{Koyama:2015vza}.

In light of such tight constraints, viable MG candidates usually invoke a restoring ``screening" mechanism \cite{Khoury:2010xi,Khoury:2013tda}, which suppresses fifth forces in high density environments and allows them to comfortably pass solar system tests. The rich spectrum of screened MG models, is often categorized into three wide classes, that reflect common qualitative features of the screening mechanism: the Chameleon models \cite{PhysRevD.69.044026, PhysRevLett.93.171104}, in which fifth forces by massive scalar fields are heavily Yukawa-suppressed when high Newtonian potentials are experienced and also the similar, but symmetry breaking, ``symmetrons" \cite{Olive:2007aj,PhysRevLett.104.231301} in which the couplings to matter are additionally weakened in dense environments. In the other two classes, the Kinetic/``K-Mouflage" \cite{Babichev:2009ee,Dvali:2010jz} and the Vainshtein mechanisms \cite{VAINSHTEIN1972393}, the non-linearities of the Lagrangian become significant in high densities, reducing the effective couplings to matter and thus recovering GR. 

The large-scale structure (LSS) of the universe is a sensitive probe of fundamental physics and as a result it can be used to infer the nature of the underlying gravitational theory. For this reason, a large set of surveys both already active, like eBOSS \citep{Comparat:2012hz} or the Dark Energy Survey (DES) \citep{Abbott:2005b} and also about to start operating in the following 10 years, like LSST \cite{Abell:2009aa}, DESI \cite{Levi:2013gra} or Euclid \cite{Laureijs:2011gra}, will study the LSS at unprecedented accuracy, offering the opportunity to test the properties of gravity and gain invaluable information about the nature of the underlying mechanism responsible for the cosmic acceleration. To maximize the impact of such observational efforts, intense theoretical work is being performed, with the aim of predicting cosmological signatures for  $\Lambda$CDM as well as for all the alternative scenarios. This is performed through a combination of analytical \cite{Zeldovich:1969sb,Bouchet:1994xp}, semi-analytical \cite{Tassev:2013pn,Valogiannis:2016ane} and numerical tools \cite{Winther:2015wla}. 

Screening, while essential for a  mechanism's viability, also greatly suppresses the modified gravity signals in the high density regions, which dominate the power spectrum signal, making their detection particularly challenging even for the ambitious future surveys of the LSS. This has motivated the consideration of density transformations that up-weight the lower density regime in favor of the higher, screened densities, so as to enhance MG signals in a density-dependent way. Penalizing the higher densities to increase the amount of information encoded in the 2-point statistics of cosmological density fields has been a valuable strategy even in the context of $\Lambda$CDM considerations. A logarithmic transform of the density field \cite{2009ApJ...698L..90N,Wang:2011fj,2012PhRvL.108g1301C}, makes the field more Gaussian allowing the recovery of more information from the 2-point function. Clipping the very high densities \cite{2011PhRvL.107A1301S,Simpson:2013nja} has also been  found to produce similar beneficial effects.
In the context of MG, clipping the screened densities \cite{Lombriser:2015axa} allows better discrimination between MG and GR, while in \cite{Llinares:2017ykn} a new generalized restricted logarithmic transform was found to boost signals. 

In this work, we investigate the performance of a new density transformation that up-weights the significance of lower densities and was first proposed in \cite{White:2016yhs}, both as a simple density transformation and as a marked correlation function. We find such a function to provide discriminatory power between MG and GR and to increase the Fisher information significantly. Furthermore, we perform a systematic comparison of the performance of these various transformations and discuss our approach in the context of related work in the literature.

The organization of the paper is as follows: in Sec. \ref{sec:Formalism} we first review the MG models studied, the simulation data used and the different density transformations considered.
In Sec. \ref{sec:Results} we present our results, assessing the performance of the different functions, before concluding and discussing future work in Sec. \ref{sec:conclusions}.
\section{Formalism}
\label{sec:Formalism}

\subsection{Modified gravity models}
\label{sec:Models}

The most general form of a Lagrangian that describes ghost-free scalar-tensor extensions to GR is of the known Horndeski form \cite{Horndeski1974,PhysRevD.84.064039}.
If by $M_{Pl}$ we denote the reduced planck mass $M_{Pl}=\frac{m_{Pl}}{\sqrt{8\pi G}}$, by R the Ricci scalar, and by ${\mathcal L}_m$ the matter sector component with fields $\psi_m$ that possess non-minimal coupling to the scalar field $\phi$, the Einstein frame form of such a Lagrangian is
\begin{equation}
\mathcal{L}=\frac{M_{Pl}^2}{2}R+\mathcal{L}(\phi,\partial_{\mu} \phi,\partial_{\mu}\partial^{\mu}\phi)+\mathcal{L}_m(e^{2\beta(\phi)\phi/M_{pl}}g_{\mu\nu},\psi_m). 
\end{equation}
The particular subclass that contains the screening mechanisms considered here, the chameleons and the phenomenologically similar symmetrons, corresponds to a scalar field Lagrangian of the form
\begin{equation}
\mathcal{L} = -\frac{1}{2} \left(\nabla\phi \right)^2-V(\phi).  
\end{equation}
The conformal coupling to matter, expressed through the dimensionless coupling constant $\beta(\phi)$, gives rise to an effective potential
\begin{equation}
V_{eff} = V(\phi) + \frac{e^{\beta\phi/M_{pl}}\rho_m}{M_{Pl}},
\end{equation}
which consists of the self-interaction potential $V(\phi)$ and a matter dependent component. The qualitative features of the particular screening mechanism are incorporated into the interplay between these two components. In the chameleon screening, $V(\phi)$ is of runaway form and in high densities the field settles down to a minimum of $V_{eff}$, becomes very massive and decouples. 
In the symmetron model, on the other hand, the interaction potential is of the ``Mexican hat" symmetry breaking form \cite{PhysRevLett.104.231301}, which additionally generates a density-dependent coupling. In low-density regions, spontaneous symmetry breaking allows coupling to matter, while in high-density environments the symmetry is restored, the coupling to matter vanishes and GR is recovered.

\subsubsection{The $f(R)$ model}
Adding a non-linear function of the Ricci scalar $R$ to the String-frame expression of the Einstein-Hilbert action, has been shown to produce cosmic acceleration, making the so-called $f(R)$ theories \cite{Carroll:2003wy} widely-studied modified gravity models. Here we consider the Hu-Sawicky $f(R)$ model \cite{Hu:2007nk}, which can be incorporated \cite{Brax:2008hh} into the chameleon screening formalism with $\beta=1/\sqrt{6}$ and is usually parametrized as
\begin{equation}
f(R)=-m^2\frac{c_1\left(R/m^2\right)^n}{c_2\left(R/m^2\right)^n+1},
\end{equation}
where $m=H_0\sqrt{\Omega_{m0}}$, is a characteristic mass scale determined by the the Hubble Constant $H_0$, and $\Omega_{m0}$, the matter fractional energy density today. The additional requirement of matching the $\Lambda$CDM background expansion dictates that $\frac{c_1}{c_2}=6\frac{\Omega_{\Lambda0}}{\Omega_{m0}}$ and the two final free parameters of the model are $\bar{f}_{R_0}=\frac{df(R)}{dR}\big|_{z=0}$ and $n$, with 
\begin{equation}
\bar{f}_{R_0} =-n\frac{c_1}{c_2^2}\left(\frac{\Omega_{m0}}{3(\Omega_{m0}+\Omega_{\Lambda0})}\right)^{n+1}.
\end{equation}
By $\Omega_{\Lambda0}$ above we denote the dark energy fractional energy density today.
Finally, within this formulation the characteristic model-dependent mass takes the form
\begin{equation}
m(a)=\frac{1}{2997}\left(\frac{1}{2|\bar{f}_{R_0}|}\right)^{\frac{1}{2}}\frac{\left(\Omega_{m0} a^{-3}+4\Omega_{\Lambda0}\right)^{1+\frac{n}{2}}}{\left(\Omega_{m0}+4\Omega_{\Lambda0}\right)^{\frac{n+1}{2}}} [Mpc/h].
\end{equation}
In this work, we will consider models that correspond to $n=1$ and $\abs{\bar{f}_{R_0}}=\{10^{-6},10^{-4}\}$, that correspond to representative choices of a weak and a strong modification choice respectively. 

\subsubsection{The symmetron model}
\label{symsec}

The free parameters of the symmetron model presented previously, are the scale factor at which symmetry breaking occurs,
 $a_{ssb}$, the force length range $\lambda_{\phi0}$ and the coupling parameter $\beta_0$. The characteristic coupling and mass take the form
\begin{equation}
\label{symscreen}
\begin{split}
\beta(a) & = \beta_0\sqrt{1-\left(\frac{a_{ssb}}{a}\right)^3} \\
m(a) & = \frac{1}{\lambda_{\phi0}}\sqrt{1-\left(\frac{a_{ssb}}{a}\right)^3} 
\end{split}
\end{equation}
We study the model with the choice of values $a_{ssb}=0.5,\lambda_{\phi 0}=1Mpc/h$ and $\beta_0=1$ for the free parameters, which represents a viable, realistic candidate based on the current experimental constraints. 
\subsubsection{N-body Simulations}
\label{sim}
In order to produce accurate realizations of the LSS for a wide range of scales, analytical considerations are inadequate due to the non-linear nature of the collapsed structures and as a result we have to resort to full blown N-body simulations. In the case of MG scenarios, the situation is further complicated by the need to accurately capture the screening effects, which are fundamentally incorporated in the non-linearities. In this paper, we use $z=0$ density snapshots that have been produced in CDM N-body simulations presented in \cite{Valogiannis:2016ane}. The simulations were performed using a suitably modified version of A. Klypin's PM code \cite{Klypin:1997sk}, in which the MG screening was captured effectively through the attachment of a phenomenological thin shell factor to the fifth force term \cite{Winther:2014cia}. The simulations were initialized at an initial redshift of $z_i=49$, for 40 random initial seeds, for a background $\Lambda$CDM cosmology that corresponds to $\Omega_{\Lambda0}=0.75$, $\Omega_{m0}=0.25$, $\sigma_8=0.8$, $n_s=1.0$ and $h=0.7$. The simulation box side L and number of particles used $N_p$ were L=200 Mpc/h and $N_p=256^3$ respectively, while the density was resolved in a $512^3$ grid using the Cloud-In-Cell (CIC) assignment scheme. More details on the specifics of the simulations and the screening implementation can be found at \cite{Valogiannis:2016ane}.

\subsection{Density transformations}
\label{sec:transformations}

The fundamental quantity of interest in our analysis, is the fractional cold dark matter over-density $\deltam(\bold{x},a)$, which is defined as
\begin{equation}
\deltam(\bold{x},a)=\frac{\rho_{m}(\bold{x},a)}{\bar{\rho}_{m}}-1,
\end{equation}
with $a$ the scale factor, $\rho_{m}(\bold{x},a)$ the matter density in each grid cell and $\bar{\rho}_{m}$ the mean density at the cosmological time considered (which is $a=1$ for our analysis). 

Using the CIC interpolation scheme, the density field, $\deltam$, from each snapshot is reconstructed on a $256^3$ resolution Cartesian grid and is then projected onto three orthogonal 2D planes that correspond to the 3 independent Cartesian axes. Through this process, the 40 random seeds initially produced for each model, generate 120 independent realizations. We find that 2D projections are significantly better for considering the transformations than the 3D density fields, sampled in the initial $256^3$ snapshots,  because the 3D cells, when up- and down- weighted with the transformation, are more sensitive to the sparse sampling and shot noise. The 3D cell size, which is an arbitrary choice, that corresponds to this choice of parameters, is equal to $0.5$ $\left(\frac{Mpc}{h}\right)^3$. This choice of cell volume was found to provide the best combination of low shot noise and high resolution.

Through a specific transformation, a new field $\deltam'=f(\deltam)$ can be constructed, with the aim of enhancing the amount of information that can be extracted. In the following section, we briefly introduce the various density transformations investigated in this paper and focus on the key aspects that will be relevant to our analysis.

\subsubsection{Logarithmic transformation}
%
The non-linearities in the dark matter power spectrum have been shown \cite{2009ApJ...698L..90N,Wang:2011fj,2012PhRvL.108g1301C} to become significantly smaller, and the amount of carried information significantly greater, in terms of signal-to-noise, when the fractional matter over-density undergoes a transformation
\begin{equation}\label{eq:log}
\deltam'=\ln \left(\deltam+1\right).
\end{equation}
Besides restoring the linear character of the power spectrum, down-weighting screened regions through a logarithmic mapping \cite{Lombriser:2015axa,Llinares:2017ykn} can serve to enhance the predicted power of MG signals.

After such a mapping is performed, a large-scale multiplicative bias, $b_{log},$ develops \cite{2017MNRAS.464L..21R} between the power spectra of the original and transformed fields, as given by
\begin{equation}\label{var}
P_{\ln(1+\deltam)}(k)=\frac{\sigma^2_{\ln(1+\deltam)}}{\sigma^2_{\delta}}P_{\deltam}(k)= b^2_{log} P_{\deltam}(k),
\end{equation}
with (\ref{var}) being valid as $k\rightarrow0$ and where $\sigma^2$ denotes the variances of the density fields, as calculated by integrating over the corresponding power spectra. Predicting the developed multiplicative bias, as has been also performed through other expressions proposed in \cite{2009ApJ...698L..90N, Wolk:2015upa}, will be particularly useful in interpreting the large-scale behavior of ratios of transformed fields in section \ref{sec:Results}, when compared to the ratios of standard power spectra. Measuring ratios of transformed-density power spectra could, however, be performed directly, without requiring knowledge of the ordinary power spectrum or biases with
respect to it. It should be noted here, that we calculate $\sigma^2_{\ln(1+\deltam)}$ through a direct integration of the logarithmic power spectrum over a top-hat filter, rather than making use of the phenomenological formula proposed in \cite{2017MNRAS.464L..21R}.

\subsubsection{Clipped  transformation}

Another transformation that reduces the contribution of higher over-densities, for the sake of maximizing the extraction of cosmological information from the up-weighted, less dense regions, is clipping \cite{2011PhRvL.107A1301S,Simpson:2013nja,Lombriser:2015axa}. In this procedure, all densities higher than a desired threshold $\delta_0$ are truncated to form a new distribution:
\begin{equation}\label{clipstr}
\deltam'=\delta_c=  \left\{\def\arraystretch{1.2} 
  \begin{array}{@{}c@{\quad}l@{}}
   \deltam  & \text{if $ \deltam < \delta_0$ }\\
    \delta_0 & \text{if $\deltam>\delta_0.$}\\
  \end{array}\right.
\end{equation}
After (\ref{var}) is applied, the new density field is ``renormalized" using the new mean density of the distribution, in order to ensure that $\langle \delta' \rangle=0$, which was proposed in \cite{2011PhRvL.107A1301S} to get results that are rather insensitive to the choice of the threshold. In our analysis we found such a prescription to perform slightly better in terms of the Fisher information, compared to when the density field is not ``renormalized", and so this is the one adopted. Just like the logarithmic transform, clipping enhances the extracted signals not only in $\Lambda$CDM, but also in MG, by emphasizing on regions not subject to screening. 

Similar to the result of (\ref{var}), the response of the density power spectrum to clipping on linear scales can be calculated \cite{Simpson:2013nja} by 
\begin{equation}\label{varclip}
P_{c}(k)=\frac{\sigma^2_c}{\sigma^2_{\delta_c}} P_{\delta}(k) ,
\end{equation}
where $P_{c}(k)$ is the power spectrum of a clipped field $\delta_c$, and the variances are again given by integrating the power spectra. The validity of (\ref{varclip}) can, if desired, be extended into the mildly non-linear regime after the introduction of perturbative one-loop contributions \cite{Simpson:2013nja}.

 Given that the value of $\delta_0$ itself is dependent on the details of each simulation, the fraction of cells clipped is a more easily transferable descriptor and so this is the one we will report.
%
\subsubsection{Marked  transformation}
\label{section:markedsect}

The use of density-dependent marks has been explored in the context of breaking degeneracies in halo occupation distributions \cite{White:2008ii}, and as a probe of identifying MG signatures in the LSS \cite{White:2016yhs}.

In this work, we consider an analytical function as a means of re-weighting the density field,
\begin{equation}\label{markdel}
\deltam'=m(\deltam)=\left(\frac{\rho_*+1}{\rho_*+\rho_m}\right)^p=\left(\frac{\rho_*+1}{\rho_*+\bar\rho_m(\deltam+1)}\right)^{p}. 
\end{equation}
where $\rho_*$ and $p$ are free parameters and $\rho_m$ the grid cell density field, in units of the mean density $\bar{\rho}_m$. 

\section{ Results}
\label{sec:Results}

\begin{figure*}[!tb]
\bc
{\includegraphics[width=0.99\textwidth]{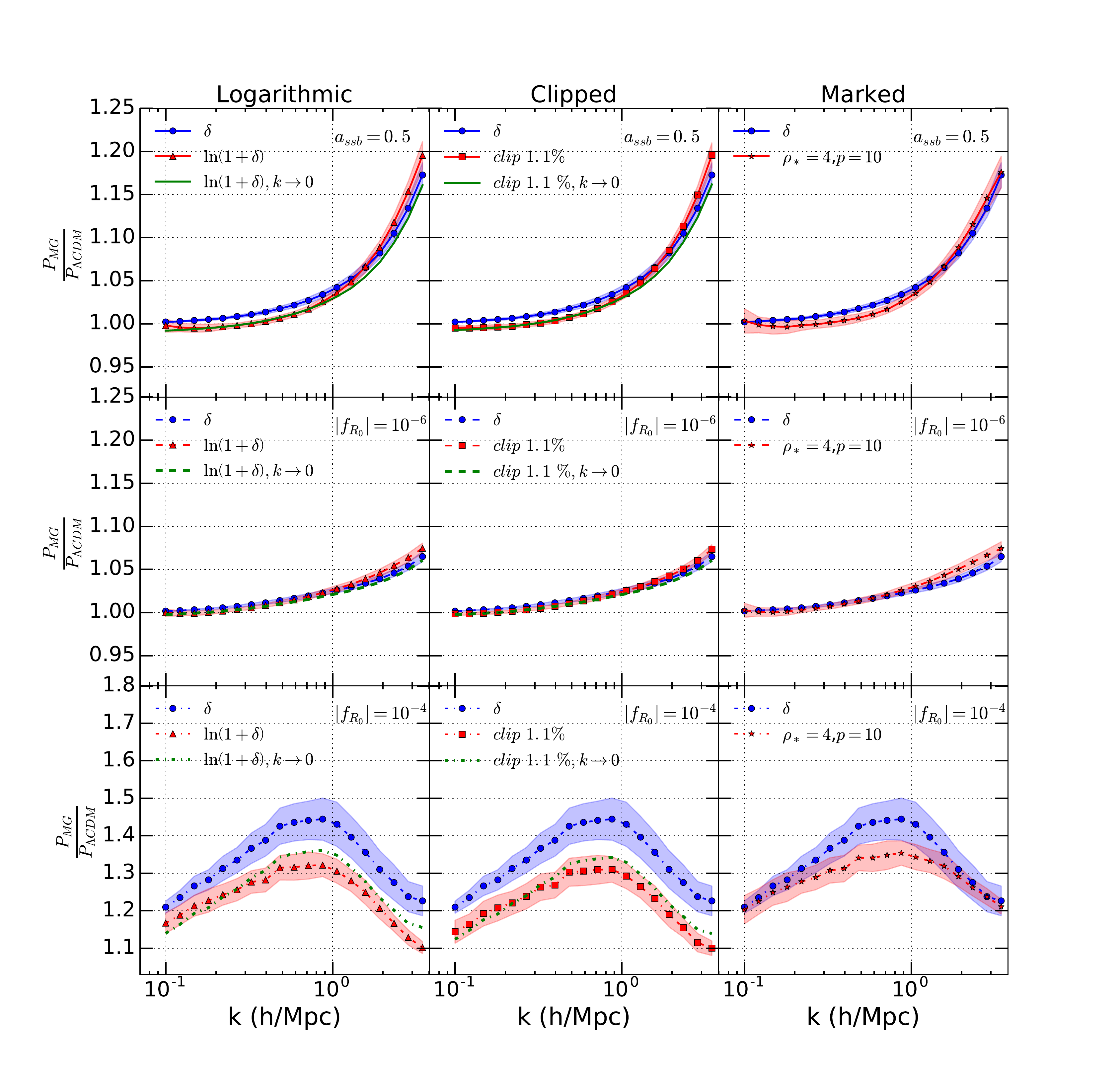}
}
\caption{A side-by-side comparison of the ratios of the MG matter power spectra, $P_{MG}$, for the symmetron [top row], $|f_{R_0}|=10^{-6}$  [middle row] and $|f_{R_0}|=10^{-4}$  [bottom row] modified gravity models relative to $\Lambda$CDM,  $P_{\Lambda CDM}$. For each model, the ratio is shown for the three density transformations:  logarithmic transformation [left column, red triangle], the clipped density field when 1.1\% of the volume is clipped [center column, red square] and the ``marked" transform, $m(\deltam$) [right column, red star], with $\rho_*=4$ and $p=10$, compared to the standard density field, $\delta$ [all panels, blue circle]. The green lines in the left and middle columns show the variance-dependent, analytic predictions for the power spectrum ratios for the large scale regime, as $k\rightarrow0$, for the logarithmic  and the clipped transformation, from equations (10) and (12) respectively. The error bands correspond to the standard deviations over the 120 realizations.}
\label{allpl}
\ec
\end{figure*}

In Figure \ref{allpl}, we present the ratios between the MG and $\Lambda$CDM matter power spectra, $\frac{P_{MG}}{P_{\Lambda CDM}}$, for all transformations considered. For the transformed fields, the ratios are found to have, in principle, different values than in the case of the standard $\delta$, on both the large and the small scales. The large scales are characterized by signal suppression with respect to the standard ratios, which, for the logarithmic and clipped transformations, is consistent with the low-k analytical predictions from equations (\ref{var}) and (\ref{varclip}). For the logarithmic case, in particular, applying (\ref{var}) twice on the individual MG and GR power spectra and dividing by parts, gives
\begin{equation}\label{ratiob}
\frac{P_{\ln(1+\deltam)}(MG)}{P_{\ln(1+\deltam)}(GR)}=\frac{b^2_{MG}}{b^2_{GR}}\frac{P_{\deltam}(MG)}{P_{\deltam}(GR)} .
\end{equation}
When $b_{MG}<b_{GR}$, which we found to be the case for all MG models, the different values of the multiplicative bias produced, as seen through (\ref{ratiob}), result in the transformed ratio being smaller at the lowest $k$ bins.  As shown in the left column of Figure \ref{allpl} for all 3 gravity models, when applied on our simulations, (\ref{ratiob}) performs well in predicting the offset between the two ratios at the smallest $k$ modes. We note that in some previous analyses, e.g.~\cite{Lombriser:2015axa}, the power spectrum ratios were normalized  applying arbitrary multiplicative factors to align the transformed ratios with unity at the lowest $k$ bins, however this is not necessary, since there is a clear analytic reason, in (\ref{ratiob}), for an inequality between the two ratios. We find that the clipped statistic has a lower standard error, in particular at small $k$, than the logarithmic and marked cases.  This can be attributed to the fact that  the clipped mapping only alters a small fraction  ($\sim$1\%) of the highest density regions, while the logarithmic and marked cases affect the whole volume, and upweight the most sparse regions associated with larger shot noise, as found in \cite{Simpson:2013nja}. 

On small scales, the signal is enhanced for the symmetron and the $|f_{R_0}|=10^{-6}$ models for all transformations by roughly 1 $\%$, and also for the $|f_{R_0}|=10^{-4}$ model in the marked transformation. 

We calculate the covariances  for $P_{MG}/P_\Lambda$, in Figure~\ref{allpl}, directly by considering the ratios of the statistics from the MG and $\Lambda$CDM  simulations with matching initial conditions.  The errors could, alternatively, be calculated from the covariances of the individual simulations, as  was proposed in  \cite{Llinares:2017ykn}, however, for simulations like ours in which the MG and $\Lambda$CDM simulations have the same initial conditions, one needs to factor in the cross-correlation between the two  \cite{Elandt-Johnson:1999}:
 \bea
 Var\left(\frac{P_{MG}}{P_{\Lambda CDM}}\right) &=&\left(\frac{\bar{P}_{MG}}{\bar{P}_{\Lambda CDM}}\right)^2\times
 \\ 
&& \hspace{-1.5cm} \left[\frac{Var(P_{MG})}{\bar{P}_{MG}^2}-2\frac{Cov(P_{MG},P_{\Lambda CDM})}{\bar{P}_{MG}\bar{P}_{\Lambda CDM}}+\frac{Var(P_{\Lambda CDM})}{\bar{P}_{\Lambda CDM}^2}\right]\nonumber.
 \eea

\begin{figure}[!tb]
\bc
{\includegraphics[width=0.48\textwidth]{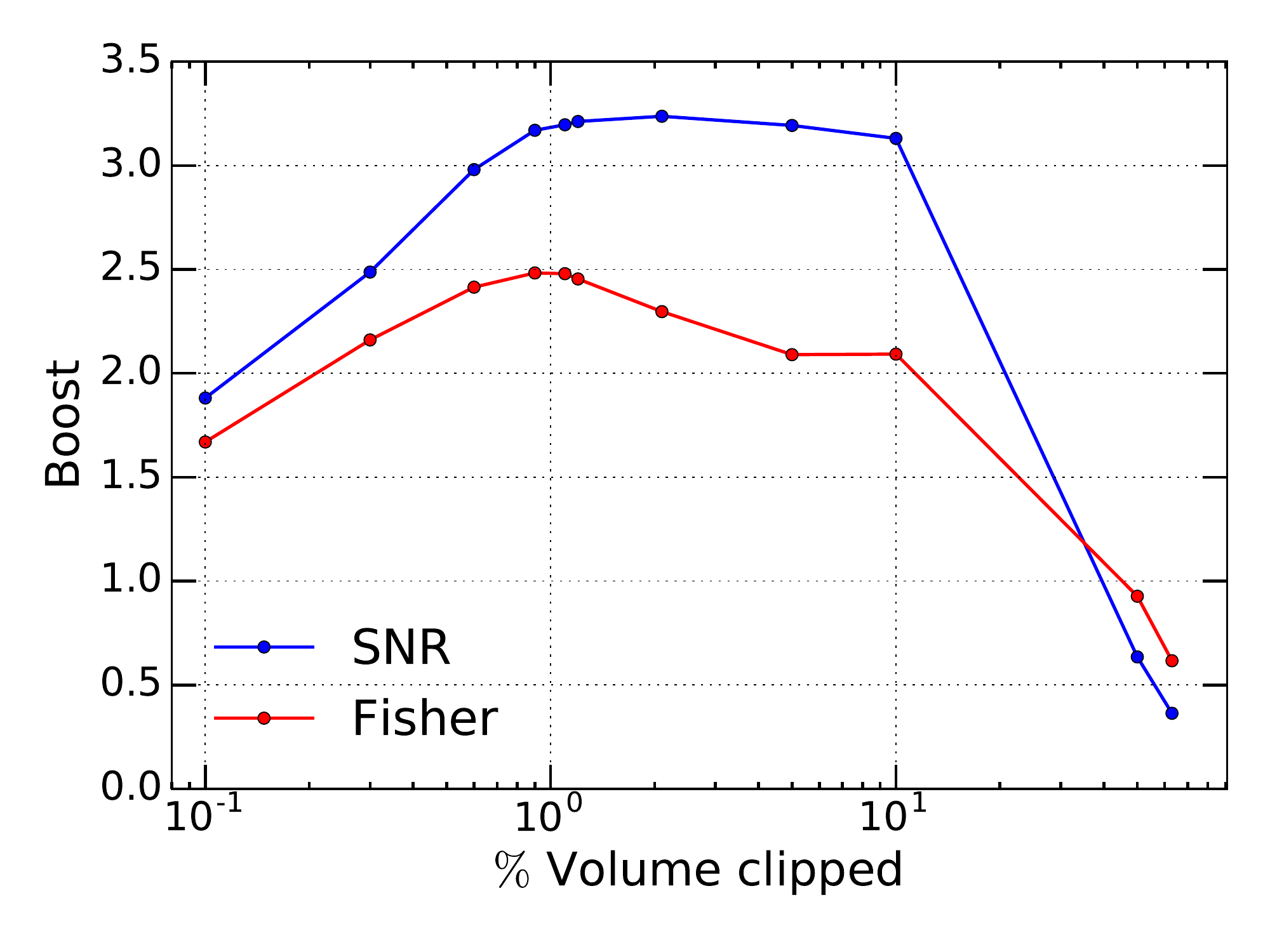}

}
\caption{The variation of the Fisher and SNR boosts, defined in (\ref{eq:Fishboost}) and (\ref{eq:SNR}), for the clipped density transformation relative to the normal density distribution, as a function of the clipping threshold, shown as the $\%$ of the  simulation volume that is clipped (i.e. has $\delta>\delta_0$). The Fisher boost is shown for an $|f_{R_0}|=10^{-6}$ cosmology, while the SNR for the $\Lambda$CDM scenario.} 
\label{clipsnr}
\ec
\end{figure}

In order to assess each transformation's efficiency in enhancing the information carried in MG signals, we calculate the matter power spectra of the 2D projected density fields, and density transformations,  as described in sec. \ref{sec:transformations} for each of the 120 independent realizations. In addition to the fractional boosts in the calculated power, expressed through the ratio $\frac{P_{MG}}{P_{\Lambda CDM}}$, the fundamental quantity of interest for statistically distinguishing MG models is the Fisher information about parameters $\alpha, \beta$ \cite{Tegmark:1996bz} :
\begin{equation}
F_{\alpha \beta}=-\left\langle \frac{\partial ^2\ln \mathcal{L}\left(data|\alpha,\beta,priors \right)}{\partial \alpha \partial \beta} \right\rangle. 
\end{equation}
Given a set of data with dependence on the parameters $\alpha, \beta$, $\mathcal{L}$ is defined as the likelihood function of the parameters from the data, and in the case of a single parameter $\alpha$, the above reduces to the Fisher information about a parameter $\alpha$:
\begin{equation}\label{firstfis}
I_{\alpha}=-\left\langle \frac{\partial ^2\ln \mathcal{L}\left(data|\alpha \right)}{\partial^2 \alpha} \right\rangle. 
\end{equation}
When restricting our focus on information encoded in the power spectra, as relevant for our analysis, (\ref{firstfis}) takes the form
\begin{equation}\label{secfis}
I_{\alpha}=-\left\langle \frac{\partial P(k_i)}{\partial \alpha}\frac{\partial ^2\ln \mathcal{L}}{\partial P(k_i) \partial P(k_j)} \frac{\partial P(k_j)}{\partial \alpha}\right\rangle, 
\end{equation}
in which the expectation value of the middle derivative term , basically the power-spectra Fisher matrix, can be well approximated \cite{Rimes:2005xs,Rimes:2005dz,Neyrinck:2006zi} by the inverse covariance $C^{-1}_{ij}$, with
\begin{equation}
C_{ij}=\frac{1}{N_{seed}-1}\sum_{r}^{N_{seed}}\left(P_r(k_i)-\bar{P}(k_i)\right)\left(P_r(k_j)-\bar{P}(k_j)\right),
\end{equation}
for the $N_{seed}=120$ realizations. It should be noted at this point that the precision in the covariance matrix calculation could be improved by applying a set of sinusoidal weightings that depend on combinations of the fundamental modes \cite{2006MNRAS.371.1188H}, as e.g. performed in \cite{2009ApJ...698L..90N}, but we do not apply such an improvement in this paper. Under these assumptions, the Fisher information about a parameter $\alpha$ takes the common form:
\begin{equation}\label{secfis}
I_{\alpha}=\sum_{i,j}^{N_{bins}}\frac{\partial P(k_i)}{\partial \alpha}C^{-1}_{ij} \frac{\partial P(k_j)}{\partial \alpha}. 
\end{equation}
In this parametrization, changes in the gravitational model are reflected upon the different values taken by the single parameter $\alpha$ (not to be confused with the scale factor $a$), which for the f(R) models is set equal to the respective values of $|f_{R_0}|$, for the symmetron equal to $z_{ssb}=1$, and equal to $0$ for $\Lambda$CDM, as the limit of both parameters that recovers GR. The numerator in the derivative terms is of course given by the difference between the corresponding MG and $\Lambda$CDM power spectra. In the general and realistic treatment involving multiple cosmological parameters, the inverse Fisher matrix is associated with the marginalized errors in the parameter estimates, while, in our single-parameter case, the unmarginalized error in the parameter estimation is predicted to be \cite{Tegmark:1996bz}, $\sigma_{\alpha}=I_{\alpha}^{-1/2}$.

To express the additional Fisher information encoded in each density mapping,  we define the ``Fisher boost", given by the ratio of $\sqrt{I_{\alpha}}$ calculated for a given mapping  to the $\sqrt{I_{\alpha}}$ by the standard density for the same cosmological model:
\bea\label{eq:Fishboost}
\mathrm{Fisher  \ boost} =\sqrt{ \frac{I_{\alpha}(\deltam')}{I_{\alpha}( \deltam)}}.
\eea

While the Fisher information provides a way to quantify the sensitivity of an estimator to changes in cosmological model parameters, the
``signal-to-noise" ratio (SNR), 
:
\begin{equation}\label{snrboost}
SNR=\sqrt{\sum_{i,j}^{N_{bins}}\bar{P}(k_i)C^{-1}_{ij}\bar{P}(k_j)},
\end{equation}
is another method used in the literature to compare  the performance of different statistics for the same cosmological model \cite{Takahashi:2009bq,2009ApJ...698L..90N}. As a comparison we consider how the SNR is affected by the choice of density transformation for the $\Lambda$CDM model.
In the same vein as in (\ref{eq:Fishboost}), we consider the change in the  $\Lambda$CDM SNR created by each transformation, as the ``SNR boost":
\bea\label{eq:SNR}
\mathrm{SNR  \ boost} = \frac{\mathrm{SNR}(\deltam')}{\mathrm{SNR}( \deltam)}.
\eea

For the clipped density transformation, the threshold, $\delta_0$, relating the fraction of the volume which is clipped, is a free parameter. In Figure~\ref{clipsnr} we show the sensitivity of the Fisher boost as $\alpha$ varies from $0$ to $|f_{R_0}|=10^{-6}$, as well as for the SNR boost, for $\Lambda$CDM, to the choice of $\delta_0$. We find that a threshold value that corresponds to clipping the $1.1\%$ most dense cells of the simulated volume maximizes both quantities. We note that our choice of using the same clipping threshold (the same value of $\delta_0$) for both MG and $\Lambda$CDM models is different than in Ref.~\cite{Lombriser:2015axa}, in which {\it different} clipping thresholds were chosen for $\Lambda$CDM and MG. The choice in \cite{Lombriser:2015axa} was made to  match the MG to $\Lambda$CDM power spectra ratios of the transformed field to those for the normal density field  at the lowest $k$ bins. As we discussed for the logarithmic case, however, the ratios of the normal and clipped density fields, in general, will not be equal on large scales but instead are determined by the variances of the original and remapped fields, through (\ref{varclip}). 

\begin{figure}[!tb]
\bc
{\includegraphics[width=0.5\textwidth]{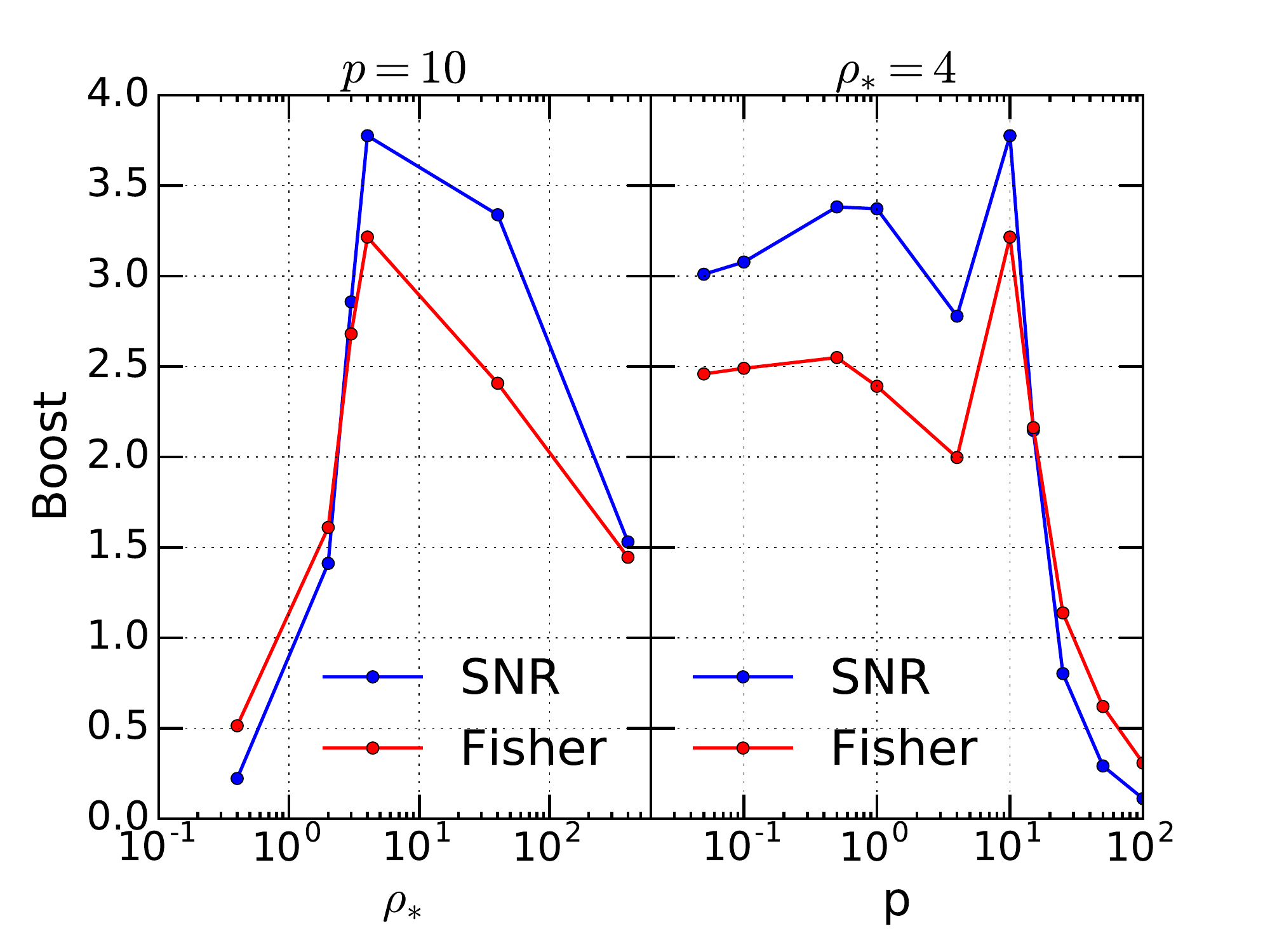}

}
\caption{The variation of the Fisher and SNR boosts, defined in (\ref{eq:Fishboost}) and (\ref{eq:SNR}),  for the marked transformation in an $|f_{R_0}|=10^{-6}$ and $\Lambda$CDM scenario, correspondingly, as a function of [left]  $\rho_*$, with fixed $p=10$, and  [right]  $p$, with fixed $\rho_*=4$.}
\label{marksnr}
\ec
\end{figure}

For the marked function, varying the values of the two free parameters, $p$ and $\rho_*$, is found to have, qualitatively, little effect on the shape and form of the transformed ratio, but a significant impact on the magnitude of the Fisher and SNR boosts, as shown in Figure~\ref{marksnr}. By fixing $p$ and varying $\rho_*$ and vice versa, we found the pair of values $p=10,\rho_*=4$ to be the optimal choice that maximizes the Fisher information as $\alpha$ goes from $0$ to $|f_{R_0}|=10^{-6}$, and the SNR boost in our GR simulations.

\begin{figure}[!tb]
\bc
{\includegraphics[width=0.48\textwidth]{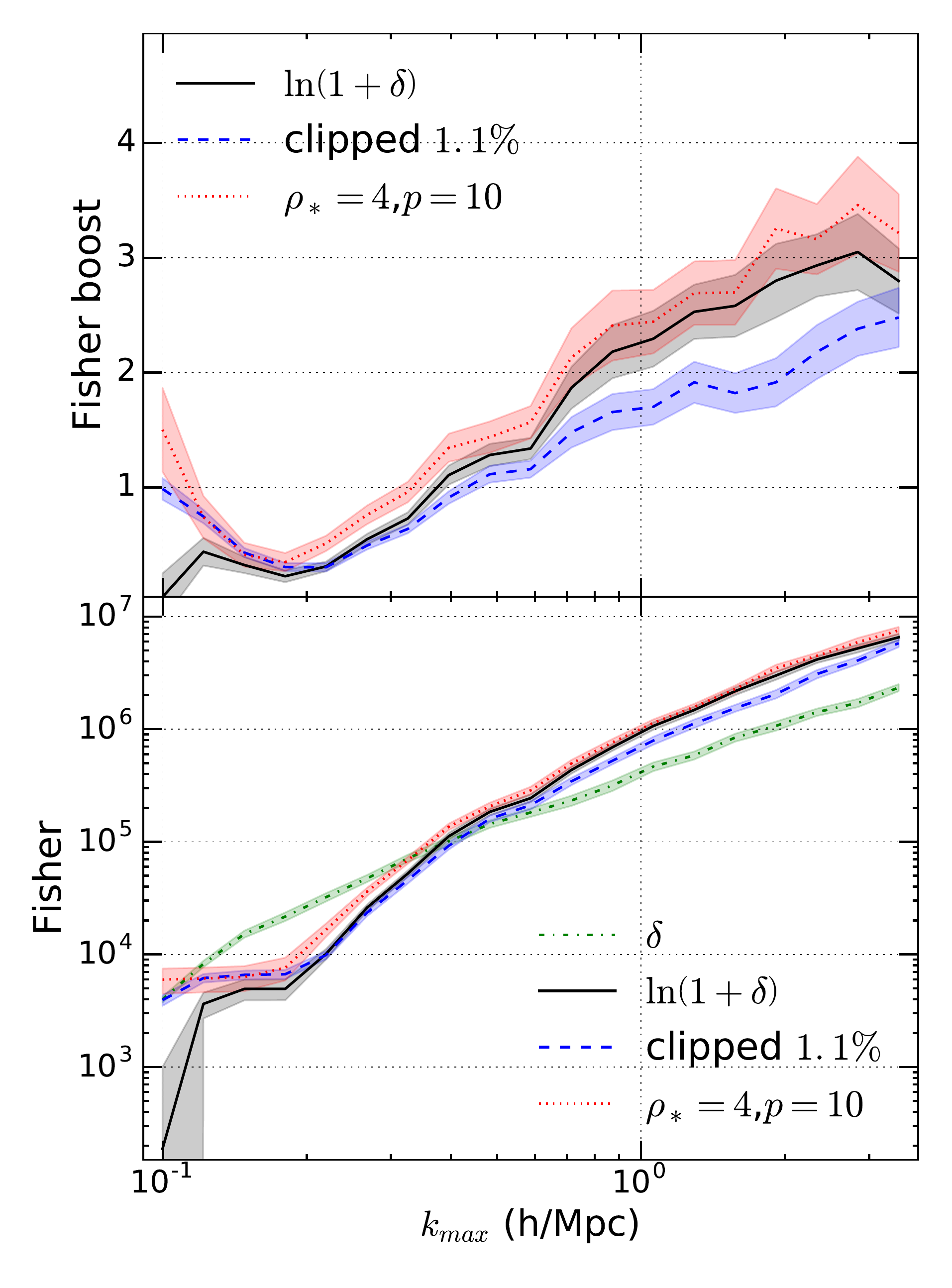}

}
\caption{[Top] The variation of the Fisher boost for the logarithmic [black, full], clipped [blue, dashed] and marked [red,dotted] transformation as a function of $k_{max}$ for the $|f_{R_0}|=10^{-6}$ case. [Bottom] The variation of the square root of the Fisher information for the standard [green, dashdot], logarithmic [black, full], clipped [blue, dashed] and marked [red,dotted] density transformation as a function of $k_{max}$ for the $|f_{R_0}|=10^{-6}$ model. The error bars have been obtained using the Jackknife approach.}
\label{fisexplain}
\ec
\end{figure}
\begin{figure}[!tb]
\bc
{\includegraphics[width=0.48\textwidth]{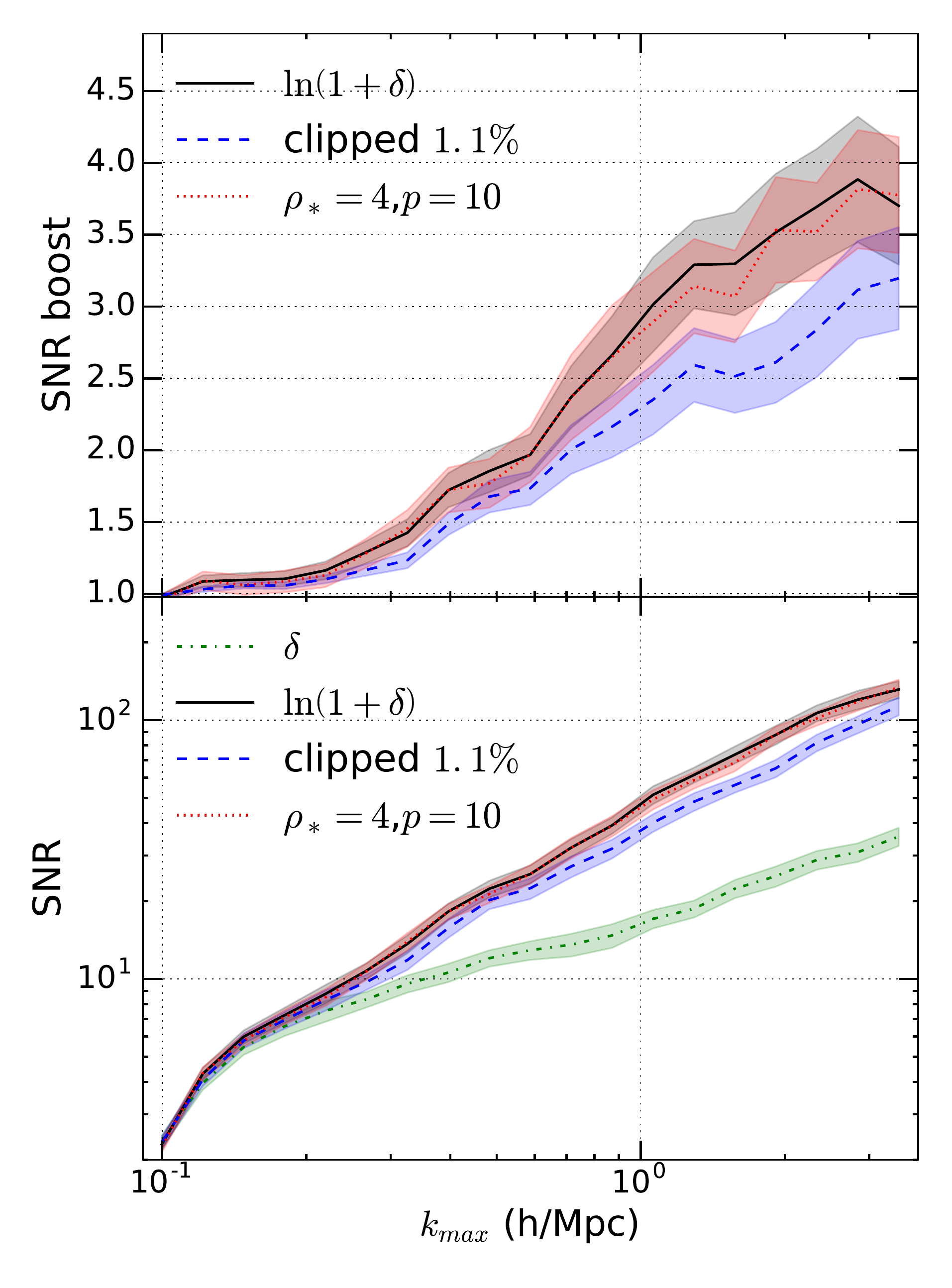}

}
\caption{[Top] The variation of the signal-to-noise ratio (SNR) boost  for the logarithmic [black, full], clipped [blue, dashed] and marked [red,dotted] transformation as a function of $k_{max}$ for the $\Lambda$CDM case. [Bottom] The variation of the signal-to-noise ratio (SNR) for the standard [green, dashdot], logarithmic [black, full], clipped [blue, dashed] and marked [red,dotted] density transformation as a function of $k_{max}$ for $\Lambda$CDM. The error bars have been obtained using the Jackknife approach.}
\label{snrexplain}
\ec
\end{figure}

As was shown in Figure~\ref{allpl},  the difference between the signal amplitudes and covariances, for the transformed statistics, relative to the normal density field is scale, as well as model, dependent. In Figure \ref{fisexplain}, the variation of the square root of the Fisher information in the power spectrum from $\Lambda$CDM to $|f_{R_0}|=10^{-6}$  is plotted as a function of the maximum wavenumber $k_{max}$, demonstrating a monotonically increasing behavior for all 4 transforms. Out of all the MG models considered, $|f_{R_0}|=10^{-6}$ represents the most viable, smallest perturbation around $\Lambda$CDM, which motivates its use as the representative example for the behavior of the Fisher information in Figures \ref{clipsnr}, \ref{marksnr} and \ref{fisexplain}. 
 When focusing our analysis on wave-modes larger than $0.4$ $h/Mpc$, all transformations comfortably predict boosts in the Fisher information, with the marked and logarithmic mappings performing better than the clipped transformation for all scales. Even though the mark predicts a higher Fisher boost than the logarithmic transformation, the predicted difference is smaller than respective error bars, making it thus hard to differentiate confidently between these two mappings with the current number of realizations. The error bars have been calculated using the Jackknife method.

In Figure \ref{snrexplain}, and in a similar manner as in Figure \ref{fisexplain}, we plot the variation in the cumulative SNR for all transformations, recovering the same qualitative behavior as in the Fisher information case. The marked and logarithmic transformations perform comfortably better, in terms of the SNR boost, than the clipping case, with the difference from each other being once again smaller than the error bars. As in the Fisher case, the error bars were obtained by the Jackknife approach.

The Fisher boosts for each of the transformations, and each modified gravity model, when calculated to three different maximum wavenumbers, 1, 1.9 and 3.5 $h/Mpc$, are summarized in Table~\ref{tab1}.

         \begin{table*}[t!]
    	\begin{tabular}{ | p{8.2em} | C{3.5em} |C{3.5em} |C{3.5em} ||  | C{3.5em} |C{3.5em} |C{3.5em} || C{3.5em} |C{3.5em} |C{3.5em} |  }
    		\cline{2-10}
		 \multicolumn{1}{c}{}& \multicolumn{9}{|c|}{Fisher Boost} 
		\\ \hline
		$k_{max}$& \multicolumn{3}{c||}{1.0 $h/Mpc$} & \multicolumn{3}{c||}{1.9 $h/Mpc$}& \multicolumn{3}{c|
		}{3.5 $h/Mpc$}
		\\ \hline
    		Transformation &  $\ln (1+\deltam)$ &$\delta_{c}$ & $ m(\delta)$&  $\ln (1+\deltam)$ &$\delta_{c}$ & $ m(\delta)$ &  $\ln (1+\deltam)$ &$\delta_{c}$ & $ m(\delta)$ 		
		 \\ \hline\hline
		Symmmetron
		& 1.6 & 1.3 & 1.5
		& 2.2 & 1.6 & 2.4 
		& 2.4 & 2.3 & 2.5	
		\\ \hline
		$f(R): |f_{R_0}|=10^{-6}$
		& 2.3 & 1.7 & 2.4
		& 2.8 & 1.9 & 3.3
		& 2.8 & 2.5 & 3.2
		\\ \hline
		$f(R): |f_{R_0}|=10^{-4}$
		& 2.1 & 1.6 &  2.2
		& 2.7 & 1.9 & 3.2
		& 2.9 & 2.2 & 3.8
		\\ \hline
		\end{tabular}
			\caption{A summary of the boost in the Fisher information when using the power spectra of the transformed, logarithmic, clipped and marked, density statistics relative to that of the standard density field for $\Lambda$CDM and the three modified gravity, symmetron and $f(R)$, models. The sensitivity of the Fisher boost to the maximum wavenumber considered is shown through the comparison of results with three different values of $k_{max}$. 
			}
    	\label{tab1}
	\end{table*}

Just like in the $|f_{R_0}|=10^{-6}$ model, the behavior of which has been shown in detail in Figure~\ref{fisexplain}, the marked and logarithmic transformations produce the highest increase in the Fisher information for the rest two gravity models under consideration, and for all $3$ wavenumbers reported, with the differences between the two mappings being smaller than the corresponding uncertainties. Furthermore, again in a similar manner with the $|f_{R_0}|=10^{-6}$ case, the Fisher boosts achieved by the optimal clipping transformation are lower than the ones produced by the other two transformations, demonstrating an overall consistent behavior for all 3 MG models. Our results also reconfirm that clipping high-density screened regions results in unscreening and enhancement of MG deviations \cite{Lombriser:2015axa} and quantifies the boost on the Fisher information, to our knowledge, for the first time.

Our results are also consistent with other studies \cite{Llinares:2017ykn}, that have shown, within the context of MG, that logarithmic mappings improve the total SNR, as they also do for $\Lambda$CDM \cite{2009ApJ...698L..90N}, and demonstrate that such transformations can be valuable in providing additional discriminatory power for difficult to detect MG models. It should be noted here that the ``restricted" logarithmic function proposed in \cite{Llinares:2017ykn}, which is essentially the the ``sliced correlation function" at fixed $\delta$ proposed in \cite{2016arXiv161006215N}, is found to produce Fisher and SNR boosts of 1.1 and 5.7 in the case of the $|f_{R_0}|=10^{-6}$ and $\Lambda$CDM models, respectively, as opposed to corresponding boosts of 3.2 and 3.8 produced by the marked transformation, for $k>3.5 h/Mpc$.

\begin{figure*}[!tb]
\bc
{
\includegraphics[width=0.48\textwidth]{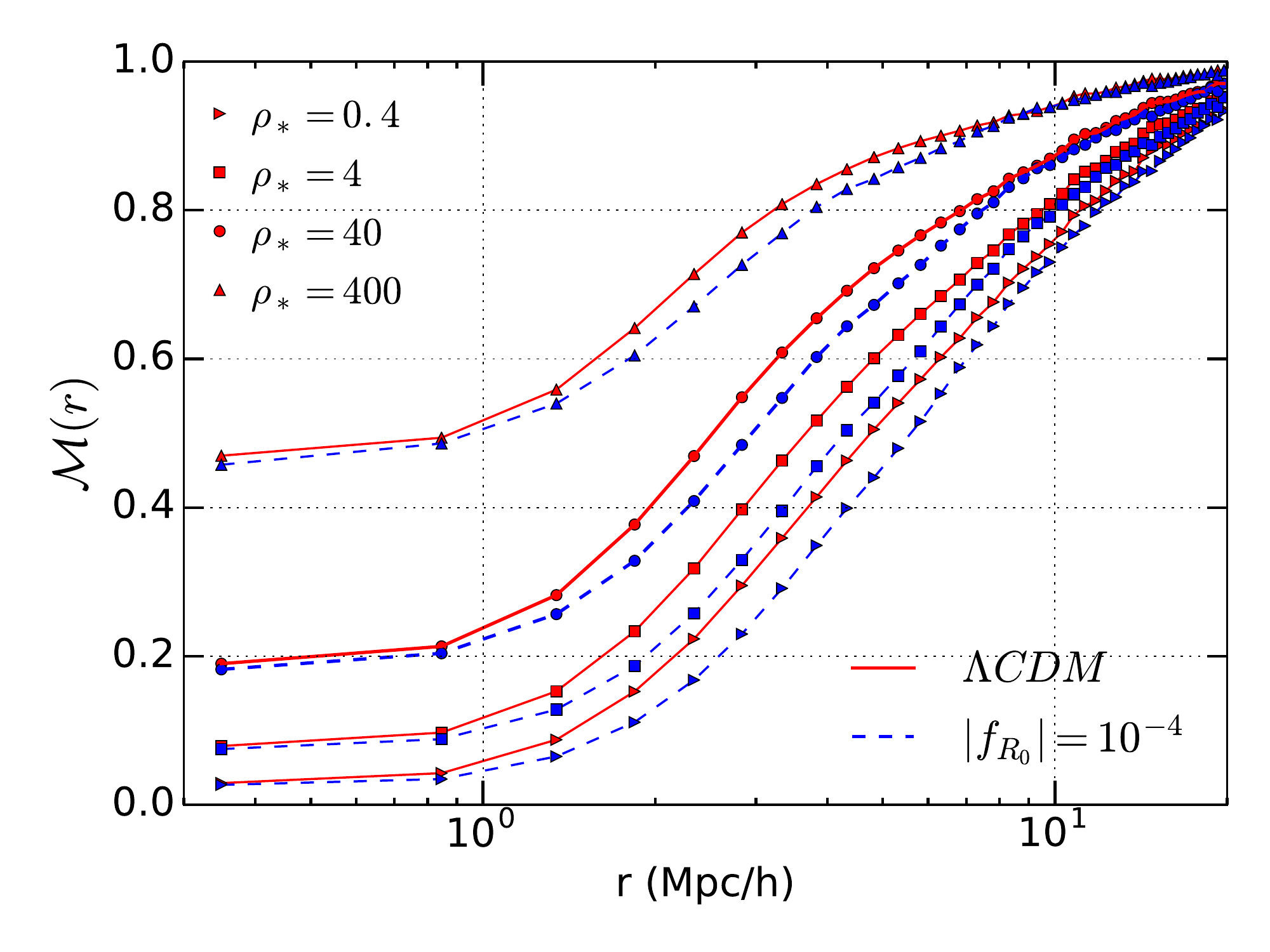}
\includegraphics[width=0.48\textwidth]{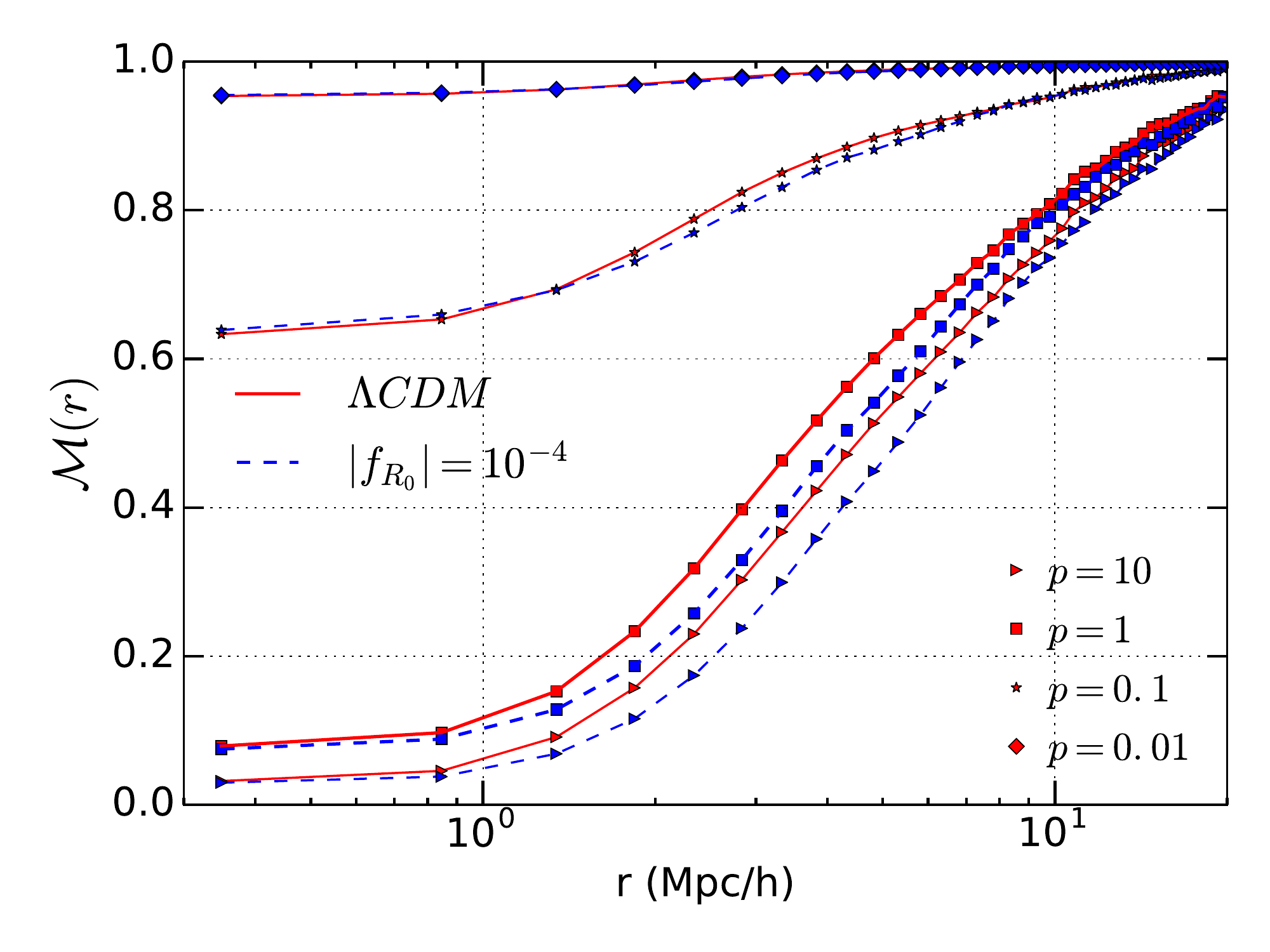}
}
\caption{The variation of the marked correlation function, $\mathcal{M}(r)$ for one realization, as a function of comoving length scale, $r$, for the $\Lambda$CDM [red, full line] and $|f_{R_0}|=10^{-4}$  [blue, dashed line]  scenario,  as a function of [left]  $\rho_*$, with fixed $p=10$, and  [right]  $p$, with fixed $\rho_*=0.4$.}
\label{markpr}
\ec
\end{figure*}

\begin{figure}[!tb]
\bc
{\includegraphics[width=0.48\textwidth]{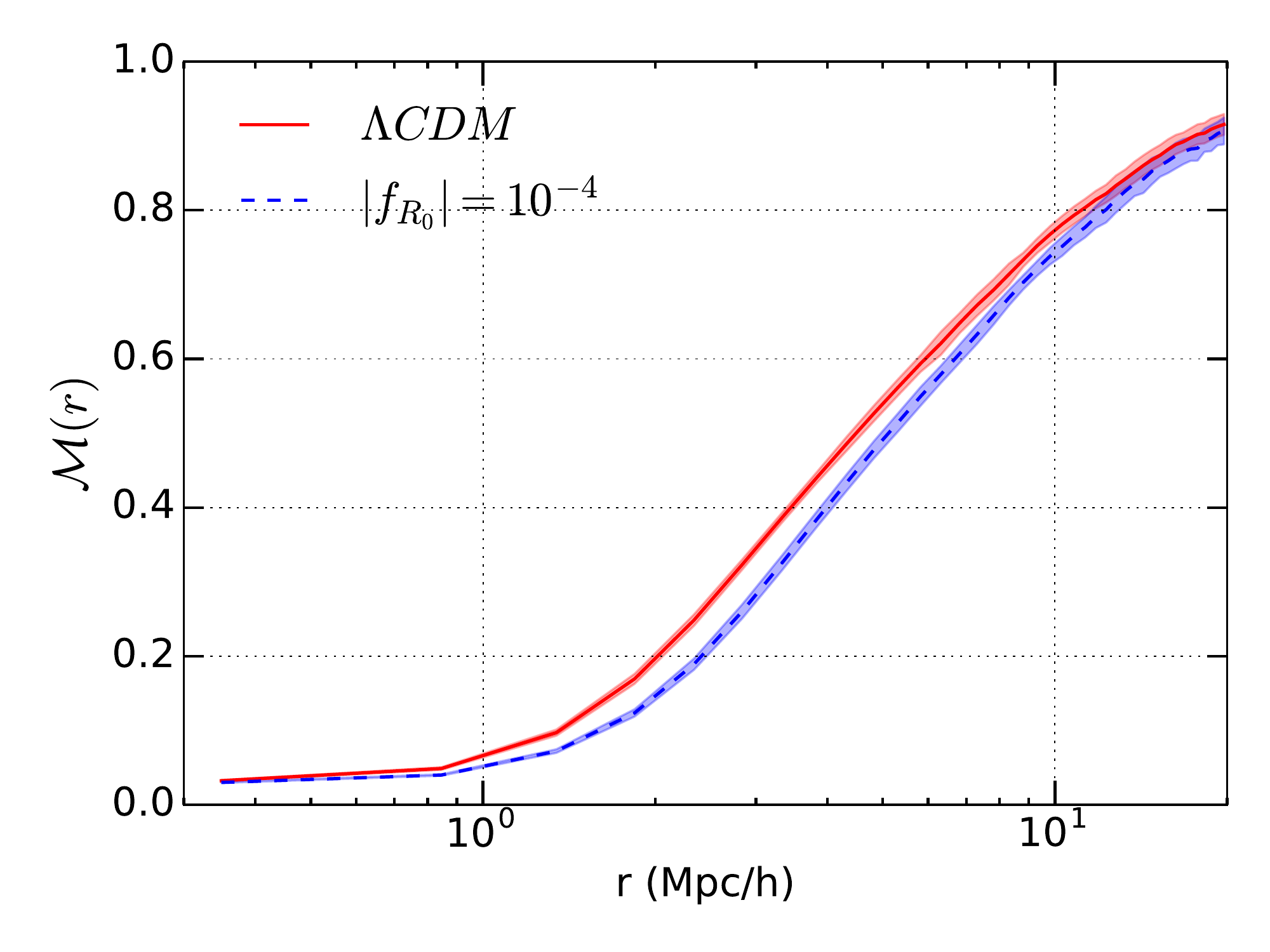}

}
\caption{The marked correlation functions, $\mathcal{M}$, for $\Lambda$CDM [red, full line] and the $|f_{R_0}|=10^{-4}$ [blue, dashed line] model, for $\rho_*=0.4$ and $p=10$, showing the average and variance over 10 realizations.}
\label{meanmark}
\ec
\end{figure}

In addition to the Fourier space statistics investigated above, we have also assessed the potential for discriminating between GR and MG models, using a real-space marked correlation function. Marked correlation functions have been proposed \cite{Beisbart:2000ja,2002LNP...600..358B,Gottloeber:2002vm,Sheth:2004vb,Sheth:2005aj,Skibba:2005kb} as an extension to the standard, autocorrelation function $\xi(r)$. We consider the marked correlation function $\mathcal{M}(r)$ of the form  \cite{White:2016yhs},
\begin{equation}\label{markedcor}
\mathcal{M}(r)=\frac{1+W(r)}{1+\xi(r)},
\end{equation}
where $W(r)$ is the correlation function weighted by the mark in (\ref{markdel}). In Figure \ref{markpr} we show, for one realization,  the variation in $\mathcal{M}(r)$, between GR and MG models  for varying values of $p$ and $\rho_*$. This demonstrates that using a marked correlation function of this form can serve as another quantity that breaks the degeneracy between MG models and the standard $\Lambda$CDM cosmological scenario. In Figure~\ref{meanmark}, we plot the marked correlation function $\mathcal{M}(r)$ with $p=10$ and $\rho_*=0.4$, for $\Lambda$CDM and the $|f_{R_0}|=10^{-4}$ model, averaged over 10 random realizations. For this analysis, we used the initial 10, out of the total of 30, realizations of the 3D density snapshots resolved in the $256^3$ mesh, rather than the projected ones, while the 3D real space autocorrelation functions were calculated using the Super W of Theta (SWOT) code \cite{2012A&A...542A...5C}. We  note that, given the functional form of (\ref{markedcor}), the observed difference between the MG and $\Lambda$CDM models is smaller than that for the standard power spectra. At $r=1.81Mpc/h$, the fractional difference  is maximal, $\mathcal{M}_{\Lambda CDM}/\mathcal{M}_{MG}=1.37$, while at $r=4Mpc/h$, $\mathcal{M}_{\Lambda CDM}/\mathcal{M}_{MG}=1.13$. The SNR boost between W and the standard $\xi$ is equal to $\sim3$ for $\Lambda$CDM.

\section{Conclusions }
\label{sec:conclusions}

Re-weighting cosmological density fields in order to suppress the contribution of dense, screened regions in favor of the low-density, unscreened regime, has been proposed as a recipe to improve the detectability of potential MG signatures. In this paper, we assess the performance of a new analytical function, first proposed in \cite{White:2016yhs}, both as a density re-mapping and as a real space marked correlation function and also perform a systematic comparison with the logarithmic and clipping transformations. Besides the fractional deviation in the dark matter power spectra, $\frac{P_{MG}}{P_{\Lambda CDM}}$, each transformation is assessed through the boost, with respect to the standard density field, in the Fisher information in the power spectra for all MG models, as well as through the boost in the total signal-to-noise ratio for $\Lambda$CDM.

By exploring the parameter space of the ``marked" density transformation, we found the parameter choice of $p=10$, $\rho_*=4$, to be the one that produces the maximum boost in the Fisher information for the $|f_{R_0}|=10^{-6}$ model, as well as the highest increase in the signal-to-noise in $\Lambda$CDM. The logarithmic mapping was found to perform roughly equally well, within the levels of accuracy, in maximizing these quantities, while both transformations were found to be superior to clipping of density peaks. These results, that also hold for the rest of the gravity models considered, demonstrate that the marked tracer could serve as a useful tool with which to discriminate between MG models and the standard cosmological scenario.

The value of the clipping threshold that truncates the densest $1.1 \%$ of each snapshot, was found to be the optimal one that simultaneously produces the maximum boosts in the Fisher information and the total signal-to-noise ratio, for all models considered. By studying the performance as a function of the maximum Fourier mode, $k_{max}$, included, we found clipping to predict smaller boosts compared to the other two transformations at all scales, while still performing considerably better than the standard density field.

Finally, we assessed the discriminatory potential of a real-space, marked correlation function of the form (\ref{markedcor}), which, tested on the $|f_{R_0}|=10^{-4}$ model, was found to provide a maximum difference relative to $\Lambda$CDM of $37\%$ at $r=1.81 Mpc/h$ and a $\Lambda$CDM SNR boost of $\sim$3, comparing $W$ to $\xi$, clearly demonstrating the power of such a real space statistic.

In this work, we have focused on the application of the statistics using the dark matter particle distribution from the N-body simulations. We recognize that in reality surveys sample astrophysical, biased, baryonic tracers of the dark matter distribution, and the next natural step, that we will undertake in future work, will be to investigate the utility of these statistics on mock galaxy catalogs that more accurately represent what we will observe with upcoming surveys. Other lines of improvement could be incorporating the effects of redshift-space distortions to the current analysis. 

Models that aim to explain cosmic acceleration through modifications to GR, evade strict solar system constraints through characteristic screening mechanisms which suppress deviations in high-density environments. In our paper we demonstrate how one can, through a series of simple density transformations, differentiate more confidently between $\Lambda$CDM and alternative scenarios. Such density-dependent suppressions make the detection of potential MG signatures challenging, even for future ambitious surveys of the LSS, like the LSST, Euclid and DESI. 

\section*{Acknowledgments}
We would like to thank Baojiu Li for providing helpful comments on the paper. We also wish to thank an anonymous referee for their careful reading and useful comments on this manuscript. The work of Georgios Valogiannis and Rachel Bean is supported by NASA ATP grant NNX14AH53G, NASA ROSES grant 12-EUCLID12- 0004 and DoE grant DE-SC0011838. 

\newpage
\nocite{*}
\bibliographystyle{apsrev}

\end{document}